\pdfoutput=1
\documentclass[unnumsec,webpdf,contemporary,large]{oup-authoring-template}%

\graphicspath{{Fig/}}

\theoremstyle{thmstyleone}%
\theoremstyle{thmstyletwo}%
\theoremstyle{thmstylethree}%

\begin{document}
\journaltitle{Briefings in Bioinformatics}
\DOI{DOI HERE}
\copyrightyear{2025}
\pubyear{2025}
\access{Advance Access Publication Date: Day Month Year}
\appnotes{Paper}

\firstpage{1}


\title[Short Article Title]{Learning Hierarchical Interaction for Accurate Molecular Property Prediction}

\author[1]{Huiyang Hong}
\author[1]{Xinkai Wu}
\author[1]{Hongyu Sun}
\author[1]{Chaoyang Xie}
\author[1,*]{Qi Wang}
\author[1,†]{Yuquan Li}

\authormark{Hong et al.}


\corresp[*]{Corresponding author. \href{mailto:qiwang@gzu.edu.cn}{qiwang@gzu.edu.cn}}
\corresp[†]{Corresponding author. \href{mailto:yvquan.li@gzu.edu.cn}{yvquan.li@gzu.edu.cn}}

\address[1]{\orgdiv{State Key Laboratory of Public Big Data, College of Computer Science and Technology}, \orgname{Guizhou University}, \orgaddress{\postcode{550025}, \state{Guizhou}, \country{China}}}



\received{Date}{0}{Year}
\revised{Date}{0}{Year}
\accepted{Date}{0}{Year}


\abstract{
Discovering molecules with desirable molecular properties, including ADMET (Absorption, Distribution, Metabolism, Excretion, and Toxicity) profiles, is of great importance in drug discovery. Existing approaches typically employ deep learning models, such as Graph Neural Networks (GNNs) and Transformers, to predict these molecular properties by learning from diverse chemical information. However, these models often fail to efficiently capture and utilize the hierarchical nature of molecular structures, and often lack mechanisms for effective interaction among multi-level features. To address these limitations, we propose a \textit{Hierarchical Interaction Message Passing Mechanism}, which serves as the foundation of our novel model, the \textbf{H}ierarchical \textbf{I}nteraction \textbf{M}essage Net (\textbf{HimNet}). Our method enables interaction-aware representation learning across atomic, motif, and molecular levels via hierarchical attention-guided message passing. This design allows HimNet to effectively balance global and local information, ensuring rich and task-relevant feature extraction for downstream property prediction tasks, such as Blood-Brain Barrier Permeability (BBBP). We systematically evaluate HimNet on eleven datasets, including eight widely-used MoleculeNet benchmarks and three challenging, high-value datasets for metabolic stability, malaria activity, and liver microsomal clearance, covering a broad range of pharmacologically relevant properties. Extensive experiments demonstrate that HimNet achieves the best or near-best performance in most molecular property prediction tasks. Furthermore, our method exhibits promising hierarchical interpretability, aligning well with chemical intuition on representative molecules. We believe that HimNet offers an accurate and efficient solution for molecular activity and ADMET property prediction, contributing significantly to advanced decision-making in the early stages of drug discovery.
}



\maketitle

\section{Introduction}

Accurate prediction of molecular properties remains a crucial task in small-molecule drug discovery~\cite{DiMasi2016}. Early assessment of features such as bioactivity, solubility, permeability, and toxicity enables medicinal chemists to efficiently filter out unsuitable compounds from large screening libraries, thereby reducing the need for costly and time-consuming experimental assays~\cite{Liu2018ADMET}. Although \emph{in vitro} and \emph{in vivo} experiments remain the gold standard for property validation, their low throughput and high expense limit their practicality for large-scale screening. Consequently, there has been a growing shift toward computational, data-driven strategies that can accelerate lead optimization and reduce downstream risk in drug development~\cite{Wu2018MoleculeNet}.

Moreover, graph neural networks (GNNs) have emerged as a leading computational strategy for molecular property prediction\cite{Wu2018MoleculeNet}. By representing atoms as graph nodes and bonds as edges, GNNs jointly capture local chemical environments and global topological context without reliance on manually engineered descriptors or fixed fingerprints. Seminal architectures, such as graph convolutional networks (GCN)\cite{xu2018powerful}, graph attention networks (GAT)\cite{velivckovic2018graph} and message passing neural networks (MPNN)\cite{gilmer2017neural} have consistently outperformed classical machine‐learning baselines. Subsequent refinements, including directed MPNN (D‑MPNN)\cite{yang2019analyzing}, contextual MPNN (C‑MPNN)\cite{song2021contextual}, Graphormer\cite{xiong2022pushing} and fragment‐aware GNNs\cite{wang2023fragment} have further enriched representational power by integrating edge features, bond directionality or fragment‐level encodings. In addition, hybrid models combining graph encoders with global attention mechanisms\cite{lee2023hybrid,zhao2024attention} have demonstrated broad applicability across diverse ADMET benchmarks, underscoring the maturity and generality of hierarchical GNN approaches in the field.

However, despite this progress, virtually all existing GNN variants remain confined to single‐layer information flows—either atomic or fragmental—thus failing to model dynamic, context‐dependent interactions among functional groups that critically influence molecular behavior\cite{Xie2021HiMol, Shang2022FH-GNN}. Recent chemical studies have shown that such synergistic effects are far from additive: for example, Wu \emph{et al.} introduced the substructure masking explanation (SME) framework\cite{wu2023chemistry} to attribute property variations to specific motif combinations, revealing that randomly masking paired substructures produces markedly different attribution patterns than masking each motif in isolation. Similarly, Chen \emph{et al.}\cite{chen2024photodynamic} demonstrated non‐linear enhancement of photodynamic‐therapy activity in hydrazone‐functionalized corroles only when hydrazone and N‐Boc/N‐Ts groups co‐occur, and Al‐Jaidi \emph{et al.}\cite{aljaidi2024anticancer} found that 5‐sulfanyl thiazole derivatives exhibit potent anticancer activity only with particular aromatic substituent pairs. Moreover, BRICS decomposition\cite{degen2008automated} can efficiently identify chemically meaningful motifs that correlate with target properties, yet current GNN‐based methods typically aggregate these motifs via simple pooling or static embeddings, thereby overlooking their cooperative interplay. These observations highlight an urgent need for a model that not only processes hierarchical information but also enables learnable, cross‐layer interactions among motifs and functional groups—only then can we fully capture the complex determinants of ADMET properties\cite{Xie2021HiMol, Shang2022FH-GNN}.

To address these challenges, we propose a novel Hierarchical Interaction Message Passing mechanism\cite{Xie2021HiMol, Shang2022FH-GNN} and develop a unified molecular message passing network, HimNet. In our approach, atoms, motifs, molecules, and molecular fingerprints are treated as distinct semantic layers\cite{degen2008automated, rogers2010extended}. Crucially, we introduce the Explainable Attention Interaction module\cite{Vaswani2017Attention, Gnes2020MolInterpret}, which enables learnable modeling of interactions between functional groups. This design facilitates the capture of cross‐motif cooperative mechanisms, including hydrogen bonding\cite{Jeffrey1997HydrogenBonding}, $\pi$–$\pi$ stacking\cite{Hunter1990Aromatic}, and hydrophobic effects\cite{Chandler2005Hydrophobic}. Moreover, we incorporate a Consensus Fingerprint Enhancement module\cite{rogers2010extended, Faulon2003QSAR} that identifies latent functional patterns via multi‐fingerprint similarity analysis and fusion, thereby guiding global structural optimization. Finally, a multi‐head attention fusion mechanism\cite{Vaswani2017Attention} aligns features across hierarchical levels, significantly improving both generalizability and interpretability. Extensive experiments on eleven datasets—including eight widely‐used MoleculeNet benchmarks\cite{Wu2018MoleculeNet} and three challenging, high‐value datasets for metabolic stability\cite{Hu2021MetabolicStability}, malaria activity\cite{Gurung2020MalariaDataset}, and liver microsomal clearance\cite{Smith2019LiverClearance}—demonstrate that HimNet consistently outperforms existing hierarchical GNN models\cite{Xie2021HiMol, Shang2022FH-GNN} and feature‐fusion baselines\cite{zhao2024attention}, achieving best or near‐best performance in most tasks.

\section{Methods}

This section details our proposed HimNet model from three aspects: first, the construction of hierarchical molecular graphs; second, the core mechanisms including the Hierarchical Interaction Message Passing Network (HIMPM), molecular fingerprint encoding module, and attention fusion; finally, the experimental setup and parameter configurations.

\subsection{Hierarchical Molecular Graph Construction}

To comprehensively capture molecular features and interactions at different scales, we constructed a three-level hierarchical graph representation including atom-level, motif-level, and global-level\cite{Xie2021HiMol,Shang2022FH-GNN}. This multi-scale representation simultaneously encodes local chemical details and complex topological relationships among molecular substructures.

Formally, the hierarchical molecular graph is defined as $G = (V, E)$, where the node set $V = \{V_a, V_m, V_g\}$ consists of atom nodes $V_a$, motif nodes $V_m$, and a global node $V_g$\cite{Xie2021HiMol}. The edge set $E = \{E_a, E_m, E_{am}, E_{mg}\}$ comprises atom-atom bonds $E_a$, motif-motif relationships $E_m$, atom-motif associations $E_{am}$, and motif-global connections $E_{mg}$.

The construction process begins by parsing the molecular topology from SMILES strings\cite{Weininger1988SMILES} to obtain atom nodes $V_a$ and chemical bonds $E_a$. Next, we apply the BRICS (Breaking of Retrosynthetically Interesting Chemical Substructures) decomposition algorithm\cite{degen2008automated} to identify functional groups, generating the motif node set $V_m$. Based on the decomposition results, we establish atom-motif associations $E_{am}$ and motif-motif connections $E_m$. Finally, we introduce a global node $V_g$ and establish connections $E_{mg}$ with all motif nodes, forming a complete hierarchical molecular graph.

For feature initialization, atom node features include atom type, hybridization state, formal charge, and chemical environment information\cite{Gilmer2017MPNN,Wu2018MoleculeNet}; motif node features are obtained by aggregating their constituent atoms' features; and the global node feature is formed by aggregating all motif features\cite{Xie2021HiMol}. This structured hierarchical representation lays the foundation for subsequent message passing and feature fusion.

\subsection{Hierarchical Interaction Message Passing Network (HIMPM)}

Hierarchical Interaction Message Passing Network (HIMPM) is the core innovative component of our model, redefining the traditional message passing paradigm by integrating hierarchical interaction attention mechanisms into the message propagation process, enabling the network to simultaneously model cooperative effects and potentially non-linear interactions across different structural hierarchies\cite{Gilmer2017MPNN, Vaswani2017Attention}. As shown in Fig. 1a, HIMPM consists of three organically integrated functional modules: an innovative message passing module, a multi-scale interaction module, and a global-local attention module\cite{Xie2021HiMol}.

The innovative message passing module transcends the limitations of traditional GNN message passing by adopting a dual-channel architecture that deeply integrates directed message passing with hierarchical interaction attention mechanisms\cite{yang2019analyzing, velivckovic2018graph}. In the D-MPNN base channel, messages are propagated along directed chemical bonds from source node $v$ to target node $w$\cite{yang2019analyzing}:
\begin{equation}
m_{vw}^t = \text{Message}(h_v^{t-1}, h_w^{t-1}, e_{vw})
\end{equation}

The aggregated message for node $v$ is obtained by summing the information from all adjacent edges\cite{Gilmer2017MPNN}:
\begin{equation}
m_v^t = \sum_{w \in \mathcal{N}(v)} m_{vw}^t
\end{equation}

Unlike traditional message passing paradigms, we introduce a hierarchical interaction attention channel that not only captures long-range dependencies between nodes, but more importantly, enables structure-aware information exchange across hierarchies\cite{velivckovic2018graph, Xie2021HiMol}. These two channels are not simply running in parallel, but are deeply integrated through an innovative learnable gating fusion mechanism\cite{Li2016GatedGraph}:
\begin{equation}
\mathbf{m}_v^{(l)} = \alpha_v \cdot \mathbf{m}_v^{\text{D-MPNN}} + (1-\alpha_v) \cdot \mathbf{m}_v^{\text{HIAtt}}
\end{equation}
where $\alpha_v = \sigma(W_{\text{gate}}[\mathbf{m}_v^{\text{D-MPNN}}||\mathbf{m}_v^{\text{HIAtt}}])$ is an adaptive gating parameter, and $\mathbf{m}_v^{\text{HIAtt}}$ represents the message obtained through the hierarchical interaction attention mechanism. This fusion enables the message passing process to simultaneously maintain chemical bond topological constraints and capture cross-scale information interactions\cite{Gilmer2017MPNN, Vaswani2017Attention}.

The multi-scale interaction module, as the core unit of HIMPM, systematically models three key interaction patterns through carefully designed multi-head interaction attention mechanisms\cite{Vaswani2017Attention,Xie2021HiMol}: atom-atom interactions capture local chemical environments and long-range electronic effects\cite{Hunter1990Aromatic,Chandler2005Hydrophobic}; motif-motif interactions encode cooperative actions between higher-order functional groups\cite{degen2008automated,Shang2022FH-GNN}; and atom-motif interactions integrate microscopic chemical properties with macroscopic substructures\cite{Xie2021HiMol}.

For example, the interaction between atom node $a$ and other atom nodes $i$ is:
\begin{equation}
h_a^{\text{new}} = \sum_{i \in \mathcal{N}(a)} \alpha_{ai} \cdot \text{MultiScaleInteraction}(h_a, h_i)
\end{equation}
Here, $\text{MultiScaleInteraction}$ is our specially designed interaction function that adaptively adjusts interaction patterns according to the representation spaces of nodes at different hierarchies\cite{Vaswani2017Attention,Xie2021HiMol}. 

Similarly, the bidirectional interaction between motif node $m$ and atom node $a$ is given by:
\begin{align}
h_m^{\text{new}} &= \sum_{a \in \mathcal{N}(m)} \alpha_{ma} \cdot \text{MultiScaleInteraction}(h_m, h_a) \\
h_a^{\text{new}} &= \sum_{m \in \mathcal{N}(a)} \alpha_{am} \cdot \text{MultiScaleInteraction}(h_a, h_m)
\end{align}
These formulations extend conventional attention to heterogeneous node types and dynamically adjust interaction modes based on hierarchical position\cite{velivckovic2018graph,wu2023chemistry}, enabling HIMPM to capture complex cross-scale synergistic effects.

The global-local attention module, as the highest-level component of HIMPM, achieves deep integration of global molecular representations with local structural features. This module employs a bidirectional attention mechanism that allows the global node to selectively aggregate critical information from both atom and motif levels\cite{velivckovic2018graph,Vaswani2017Attention}:
\begin{align}
H_g^{(a)} &= W_g \cdot \sum_{a \in \mathcal{N}(g)} \alpha_{ga} \cdot h_a \\
H_g^{(m)} &= W_g \cdot \sum_{m \in \mathcal{N}(g)} \alpha_{gm} \cdot h_m
\end{align}

The final global representation integrates multi-level information through a carefully designed fusion function:
\begin{equation}
H_g^{\text{final}} = \text{GL-Fusion}(\{H_g^{(a)}, H_g^{(m)}\})
\end{equation}

Through this hierarchical fusion architecture, HIMPM breaks the barriers of information exchange between layers in traditional GNNs, achieving seamless information flow from atoms to motifs to global molecular representations, greatly enhancing the model's ability to model complex molecular structures and properties\cite{Xie2021HiMol,Shang2022FH-GNN}.

\subsection{Molecular Fingerprint Encoding Module}

As shown in Fig. 1b(ii), we designed a multi‐modal molecular fingerprint encoding module that integrates five complementary molecular fingerprints: Atom Pairs (AtomPairs)\cite{Carhart1985AtomPairs}, MACCS keys\cite{Durant2002MACCS}, Morgan Bits (MorganBits)\cite{rogers2010extended}, Morgan Counts (MorganCounts)\cite{rogers2010extended}, and Pharmacophore fingerprints\cite{Schneider1992Pharmacophore}. Each fingerprint is processed through an independent encoder to obtain a fixed‐dimensional representation:
\begin{equation}
F_i = \text{Encoder}_i(FP_i)
\end{equation}

To extract common information between different fingerprint types, we calculate the cosine similarity matrix $S_{i,j}$ between fingerprint encodings\cite{Huang2008CSimilarity} and extract shared features based on similarity:
\begin{equation}
\text{CommonFeature}_{i,j} = \left(\frac{F_i + F_j}{2}\right) \odot \mathbb{I}(\text{ElementSim}_{i,j} > \tau)
\end{equation}
where $\text{ElementSim}_{i,j} = \hat{F}_i \odot \hat{F}_j$ represents element‐wise similarity of normalized vectors.

Shared features are aggregated through similarity‐weighted summation and combined with original features:
\begin{equation}
F_{\text{final}} = \phi(W_f[F_{\text{weighted}}; F_{\text{enhanced}}])
\end{equation}
where $F_{\text{weighted}} = \sum_{i=1}^5 \alpha_i \cdot F_i$ is a dynamically weighted combination of fingerprints, and $F_{\text{enhanced}}$ represents enhanced shared features\cite{rogers2010extended}.

\subsection{Attention Fusion}

As illustrated in Fig. 1b(iv), the attention fusion module integrates features from HIMPM‑generated molecular graph representations and fingerprint encoding through a multi‑head self‑attention mechanism\cite{Vaswani2017Attention}. Each attention head contains three linear transformations: Query (Q), Key (K), and Value (V), enabling feature interactions across different representation spaces:
\begin{equation}
\text{Attention}(Q, K, V) = \text{softmax}\left(\frac{QK^T}{\sqrt{d_k}}\right)V
\end{equation}

The outputs of the multi‑head attention are concatenated and linearly transformed to obtain the final fused representation:
\begin{equation}
X_{\text{fused}} = W_O[\text{head}_1; \text{head}_2; \ldots; \text{head}_n]
\end{equation}

This module achieves complementary feature fusion, effectively integrating the structural interaction information captured by HIMPM and the chemical property information encoded by molecular fingerprints, forming a comprehensive molecular representation for downstream task prediction\cite{Vaswani2017Attention}.

\subsection{Experimental Settings}

We evaluated our HimNet model on a total of eleven datasets: eight widely‐used benchmark datasets—including five classification tasks and three regression tasks (BBBP, BACE, Tox21, SIDER, ClinTox, ESOL, FreeSolv, and Lipophilicity)\cite{Wu2018MoleculeNet}—as well as three challenging, high‐value datasets introduced in this work: Malaria (antimalarial activity regression)\cite{Gurung2020MalariaDataset}, LMC (liver microsomal clearance regression)\cite{Smith2019LiverClearance}, and MetStab (metabolic stability classification)\cite{Hu2021MetabolicStability}. Table~\ref{tab:datasets} summarizes the basic information of each dataset.

\begin{table}[htbp]
\centering
\setlength{\tabcolsep}{2pt} 
\caption{Hyperparameter settings for the HimNet model}
\begin{tabular}{ll}
\hline
\textbf{Hyperparameter} & \textbf{Value} \\
\hline
Batch size & 64 \\
Learning rate & 0.0001 \\
Hidden layer dimension & 512 \\
Molecular encoder depth & 7 layers \\
Cross-attention heads & 8 \\
Fusion attention heads & 4 \\
Dropout rate & 0.1 \\
Training epochs & 100 \\
Optimizer & Adam ($\beta_1=0.9$, $\beta_2=0.999$) \\
\hline
\end{tabular}
\label{tab:hyperparams}
\end{table}

To ensure rigorous evaluation of model performance, we adopted scaffold‐based splitting with an 8:1:1 ratio for training, validation, and test sets across all datasets, following the protocol in MoleculeNet\cite{Wu2018MoleculeNet}. This splitting method ensures that test molecules have scaffold structures different from those in the training set, effectively assessing the model's generalization capability. For classification tasks, we used the ROC‐AUC metric\cite{Fawcett2006}; for regression tasks, we evaluated performance using the root mean square error (RMSE)\cite{Chai2014}.

The HimNet model was implemented using PyTorch 1.12.0\cite{Paszke2019PyTorch} and PyTorch Geometric 2.2.0\cite{Fey2019PyG}, and trained on NVIDIA GPU environments (CUDA 11.3). Molecular operations and fingerprint generation were implemented using the RDKit 2022.03.5 library\cite{Landrum2013RDKit}, with BRICS decomposition for motif identification\cite{degen2008automated} and five types of molecular fingerprints (AtomPairs\cite{Carhart1985AtomPairs}, MACCS keys\cite{Durant2002MACCS}, Morgan Bits \& Counts\cite{rogers2010extended}, and Pharmacophore\cite{Schneider1992Pharmacophore}) for feature extraction. All experiments on all datasets followed the same preprocessing pipeline and evaluation protocol to ensure fair comparison with baseline methods.

\section{Results}\label{sec3}

\subsection{Hierarchical Interaction Message Passing Mechanism}

To uncover the intricate dependencies that span multiple structural levels, HimNet incorporates a \textbf{H}ierarchical \textbf{I}nteraction \textbf{M}essage \textbf{P}assing \textbf{M}echanism (\textbf{HIMPM}). Conventional message‐passing schemes normally operate only on directly bonded atoms or on rigidly pre‐defined fragments\cite{gilmer2017neural}. In contrast, real molecular systems often feature cooperative or competitive interactions that cross covalent boundaries—think of $\pi\!-\!\pi$ stacking between adjacent aromatic rings\cite{Hunter1990Aromatic} or the way a distant polar substituent modulates a hydrophobic cluster\cite{Chandler2005Hydrophobic}. HIMPM shatters this single‐view limitation by tying together atoms, motifs, and the molecule as a whole, enabling seamless information flow across scales and faithfully capturing these non‐additive, multi‐pathway behaviors\cite{Xie2021HiMol,Shang2022FH-GNN}.

In practice, HIMPM runs two complementary pathways in parallel. One adopts a directed message‐passing strategy (akin to D--MPNN)\cite{yang2019analyzing}, using chemical bonds to precisely encode each atom’s local environment. The other leverages \textbf{hierarchical interaction attention} to erect “long‐distance bridges” across atom–atom, atom–motif, motif–motif, and motif–global links, allowing the model to directly sense and weigh the influence between disparate structural fragments\cite{Vaswani2017Attention,Xie2021HiMol}. At each iteration, a learnable gating mechanism fuses the outputs of these pathways, dynamically balancing strict chemical bond constraints with flexible, multi‐scale semantic integration\cite{Xie2021HiMol}.

This dual‐pathway fusion of message passing and attention gives HimNet deep insight into intramolecular cooperation. Whether it is adjacent aromatic rings stacking cooperatively or hydrophobic and polar groups competing over several bonds, HIMPM makes these effects explicit through its attention weights. As a result, HimNet not only achieves notable accuracy gains on tasks such as blood–brain barrier permeability, aqueous solubility, and toxicity prediction\cite{Wu2018MoleculeNet}, but also provides clear, chemically grounded rationales for its decisions—substantially enhancing both performance and interpretability\cite{Gnes2020MolInterpret}.

\begin{figure*}[htbp]
    \centering
    \includegraphics[width=0.8\linewidth]{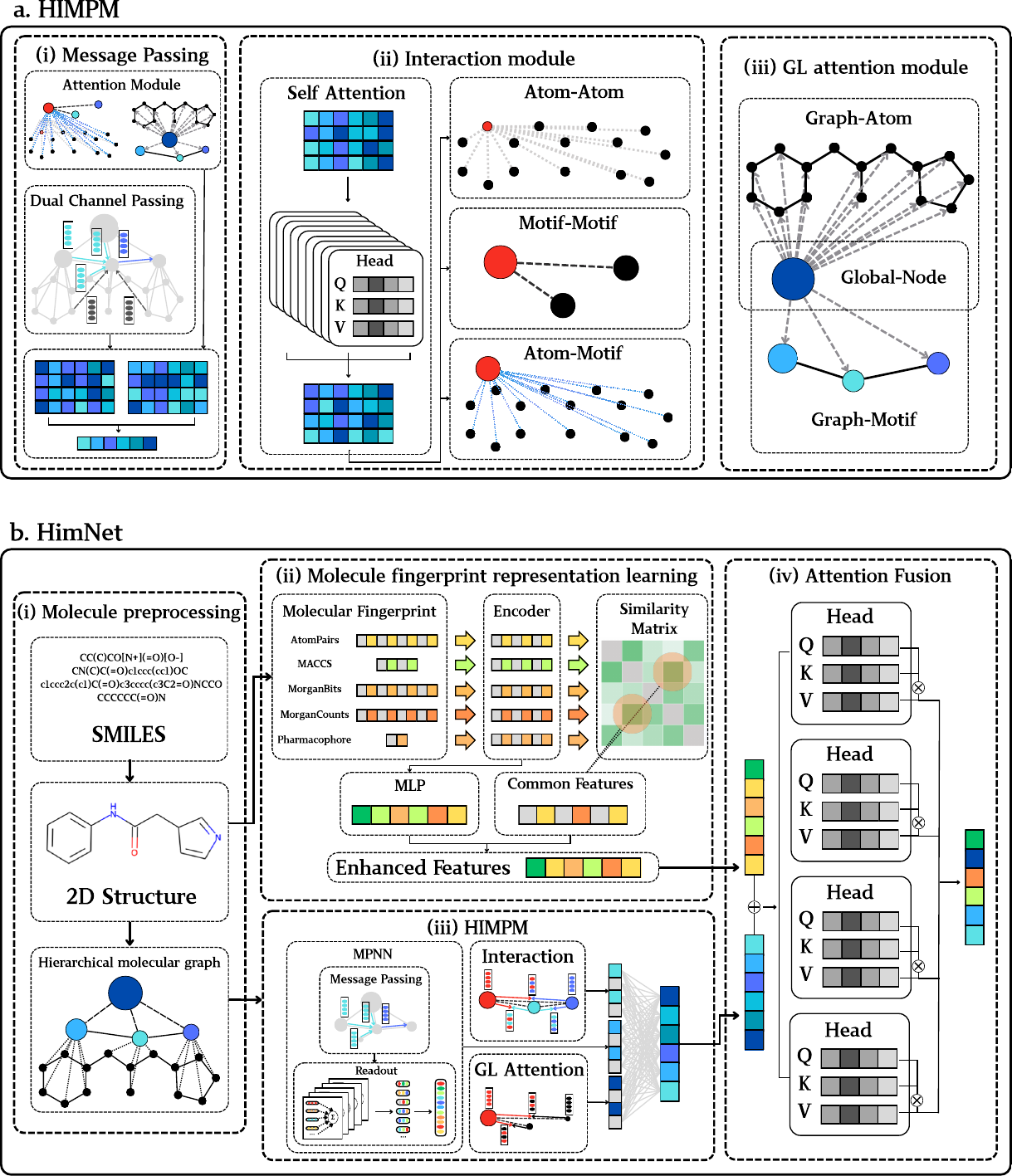}
    \caption{Overview of the proposed HimNet framework. 
(a) \textbf{Hierarchical Interaction Message Passing Module (HIMPM).} (i) Message Passing: Combines a dual-channel architecture consisting of directed message passing (D-MPNN) and a hierarchical interactive attention mechanism. (ii) Interaction Module: Models multi-scale interactions including atom–atom, motif–motif, and atom–motif relationships using cross-hierarchical attention mechanisms. (iii) Global-Local Attention Module: Aggregates information from atom-level and motif-level nodes into a unified graph-level representation through hierarchical attention fusion. 
(b) \textbf{HimNet Architecture.} (i) Molecule Preprocessing: Converts SMILES strings into 2D molecular structures and constructs hierarchical molecular graphs comprising atom, motif, and global nodes. (ii) Molecular Fingerprint Representation Learning: Encodes multiple types of molecular fingerprints and extracts common features via similarity-guided fusion strategies. (iii) HIMPM: Applies the hierarchical message passing and interaction mechanisms described in (a) to the hierarchical graph representations. (iv) Attention Fusion: Integrates multi-source features from fingerprints and hierarchical graphs using multi-head self-attention to produce expressive molecular embeddings for downstream tasks.}
    \label{fig:model_structure}
\end{figure*}

\subsection{Model Overview: HimNet}

\hyperref[fig:model_structure]{Fig.~1b} illustrates the overall architecture of HimNet, which integrates four tightly coupled stages to transform raw SMILES into a unified embedding for downstream prediction. In the first stage, \textbf{Molecule Preprocessing}, each SMILES string is parsed into a 2D chemical graph\cite{Weininger1988SMILES} and then fragmented by the BRICS algorithm\cite{degen2008automated} into atom‐level nodes, motif‐level nodes and a single global supernode. Explicit atom–motif, motif–motif and motif–global edges link these nodes, preserving fine‐grained bond topology while exposing higher‐order substructure relationships\cite{Xie2021HiMol}.

Running in parallel, \hyperref[fig:model_structure]{Fig.~1b(ii)} introduces the \textbf{Fingerprint Embedding} module, which encodes five classical molecular fingerprints—AtomPairs\cite{Carhart1985AtomPairs}, MACCS keys\cite{Durant2002MACCS}, MorganBits and MorganCounts\cite{rogers2010extended}, and Pharmacophore\cite{Schneider1992Pharmacophore}—into dense vectors. A pairwise cosine similarity matrix highlights high‐confidence shared dimensions\cite{Huang2008CSimilarity}, which are extracted as “consensus” features and then fused with the original fingerprint embeddings via learnable attention weights and a contrastive alignment loss\cite{Chen2020SimCLR}, thereby injecting global chemical priors that complement the graph encoding.

The third component is the \textbf{Hierarchical Interaction Message Passing Mechanism (HIMPM)} shown in \hyperref[fig:model_structure]{Fig.~1b(iii)}\cite{Xie2021HiMol,Shang2022FH-GNN}. One path follows directed message passing (D‐MPNN)\cite{yang2019analyzing} to rigorously propagate local atomic information along chemical bonds, while the parallel interaction‐attention path builds long‐distance bridges among atoms, motifs and the global node to capture non‐additive cooperativity such as $\pi$–$\pi$ stacking, hydrogen‐bond networks and hydrophobic clustering\cite{Vaswani2017Attention}. Learnable gating weights adaptively fuse these two channels at each iteration, ensuring strict adherence to bond constraints together with flexible integration of multi‐scale semantics.

\hyperref[fig:model_structure]{Fig.~1b(iv)} depicts the \textbf{Multi‐Head Attention Fusion} module, which aligns and merges the multi‐scale graph representations from HIMPM with the consensus fingerprint embedding through data‐driven attention weights\cite{Vaswani2017Attention}. The resulting compact, highly expressive vector drives downstream classifiers and regressors, achieving state‐of‐the‐art accuracy on blood–brain barrier permeability, solubility and toxicity prediction\cite{Wu2018MoleculeNet}. Moreover, its attention maps provide clear, chemically grounded explanations that greatly enhance interpretability\cite{Gnes2020MolInterpret}.

\subsection{Accurate Molecular Property Prediction with HimNet}\label{subsec3}

\begin{table*}[htbp]
  \centering
  \setlength{\tabcolsep}{4pt}
  \begin{tabular}{llcccc}
    \toprule
    \textbf{Data class} & \textbf{Dataset} & \textbf{\# molecules} & \textbf{\# tasks} & \textbf{Metric} & \textbf{Type} \\
    \midrule
    \multirow{3}{*}{Physicochemical}
      & ESOL (estimating aqueous solubility)                  & 1\,128 & 1  & RMSE & Regression \\
      & FreeSolv (hydration free energy)                      &   642 & 1  & RMSE & Regression \\
      & Lipophilicity (octanol–water partition coefficient LogP) & 4\,200 & 1  & RMSE & Regression \\
    \midrule
    \multirow{2}{*}{Bioactivity}
      & BACE ($\beta$‑secretase 1 inhibitor activity)               & 1\,513 & 1  & AUC  & Classification \\
      & Malaria (anti‑malarial EC\textsubscript{50})          & 9\,998 & 1  & RMSE & Regression \\
    \midrule
    \multirow{3}{*}{Toxicity}
      & Tox21 (12 toxicity endpoints)                        & 7\,831 & 12 & AUC  & Classification \\
      & SIDER (27 adverse drug reaction labels)               & 1\,427 & 27 & AUC  & Classification \\
      & ClinTox (clinical trial toxicity)                     & 1\,478 & 2  & AUC  & Classification \\
    \midrule
    \multirow{3}{*}{Pharmacokinetic}
      & BBBP (blood–brain barrier penetration)                & 2\,039 & 1  & AUC  & Classification \\
      & LMC (liver microsomal clearance in human, rat, mouse) & 8\,755 & 3  & RMSE & Regression \\
      & MetStab (metabolic stability half‑life)                & 2\,267 & 1  & AUC  & Classification \\
    \midrule
  \end{tabular}
  \caption{Comprehensive summary of the 11 benchmark datasets evaluated in this work, encompassing physicochemical, bioactivity, toxicity, and pharmacokinetic properties. The table details the number of molecules, tasks, metrics, and task types for each dataset, including both widely used MoleculeNet benchmarks and several challenging, less-explored datasets, thereby ensuring a thorough assessment of model performance and generalization.}
  \label{tab:datasets}
\end{table*}

\begin{table*}[htbp]
    \centering
    \setlength{\tabcolsep}{1pt} 
    \begin{tabular}{lcccccccc}
        \toprule
        \textbf{Dataset} & \textbf{BACE} & \textbf{BBBP} & \textbf{Tox21} & \textbf{SIDER} & \textbf{ClinTox} & \textbf{ESOL} & \textbf{Freesolv} & \textbf{Lipophilicity} \\
        \midrule
                 MolCLR        & 0.890(0.003) & 0.736(0.005) & 0.798(0.007) & 0.680(0.011) & 0.932(0.017) & 1.110(0.010) & 2.200(0.200) & 0.650(0.080) \\
        GEM           & 0.856(0.011) & 0.724(0.004) & 0.781(0.001) & 0.672(0.004) & 0.901(0.013) & 0.798(0.029) & 1.877(0.094) & 0.660(0.008) \\
        ImageMol      & 0.839(0.005) & 0.739(0.002) & 0.773(0.001) & 0.660(0.001) & 0.851(0.014) & 0.970(0.070) & 2.020(0.070) & 0.720(0.010) \\
        MolMCL$_\mathrm{GIN}$   & 0.850(0.011) & 0.741(0.006) & 0.775(0.003) & 0.667(0.008) & \textbf{0.957(0.012)} & –           & –           & –           \\
        MolMCL$_\mathrm{GPS}$   & 0.861(0.013) & 0.736(0.007) & 0.790(0.006) & \textbf{0.687(0.002)} & 0.951(0.005) & –           & –           & –           \\
        Uni-Mol       & 0.857(0.002) & 0.729(0.006) & 0.796(0.005) & 0.659(0.013) & 0.919(0.018) & 0.788(0.029) & 1.480(0.048) & \textbf{0.603(0.010)} \\
        MESPool       & 0.855(0.039) & 0.848(0.046) & 0.787(0.019) & 0.576(0.026) & 0.902(0.065) & 1.276(0.246) & 2.779(0.762) & 0.708(0.042) \\
        HimGNN        & 0.856(0.034) & 0.928(0.027) & 0.807(0.017) & 0.642(0.023) & 0.917(0.030) & 0.870(0.154) & 1.921(0.474) & 0.632(0.016) \\
        HiMol         & 0.846(0.002) & 0.732(0.008) & 0.762(0.003) & 0.625(0.003) & 0.808(0.014) & 0.833        & 2.283        & 0.708        \\
        FH-GNN$^{a}$  & 0.882(0.029) & 0.949(0.016) & 0.824(0.014) & 0.639(0.002) & 0.945(0.046) & 0.904(0.070) & 1.873(0.263) & 0.744(0.091) \\
                \textbf{HimNet}        & \textbf{0.890(0.026)} & \textbf{0.954(0.020)} & \textbf{0.826(0.018)} & 0.651(0.009) & 0.950(0.051) & \textbf{0.710(0.016)} & \textbf{1.441(0.223)} & 0.698(0.037) \\

        \midrule
    \end{tabular}
    \caption{Comparative performance of HimNet and baseline models on eight benchmark datasets. For all baselines except FH-GNN$^{a}$, results are cited from the original publications. As the original FH-GNN did not disclose its dataset splitting protocol, we reproduced FH-GNN$^{a}$ under the same scaffold-based split and experimental settings as HimNet to ensure a fair and direct comparison. }
    \label{tab:table3}
\end{table*}

\begin{table*}[htbp]
    \centering
    \begin{tabular}{lcccccc}
        \toprule
        \textbf{Dataset} & \textbf{MetStab} & \textbf{Malaria} & \textbf{LMC-H} & \textbf{LMC-R} & \textbf{LMC-M} & \textbf{LMC$_\mathrm{mean}$} \\
        \midrule
        MPNN  & 0.823(0.020)    & 1.056(0.037)    & 104.0(5.8)    & 162.9(27.4) & 125.9(20.6)   & 130.9(4.1)   \\
        GCN   & 0.696(0.055)    & 1.057(0.043)    & 103.7(5.7)    & 161.6(26.2)   & 126.4(20.2)   & 130.6(3.6)   \\
        GAT   & 0.767(0.031)    & 1.048(0.040)    & 102.6(5.5)    & 159.4(24.6)   & \textbf{124.8(17.3)}  & 128.9(3.6)   \\
        GIN   & 0.773(0.041)    & 1.045(0.038)    & 102.7(6.3)    & 158.4(22.9)   & 126.7(19.4)   & 129.3(2.2)   \\
        \textbf{HimNet} & \textbf{0.896(0.008)} & \textbf{1.029(0.030)} & \textbf{91.7(11.9)} & \textbf{109.3(0.7)} & 132.1(39.5) & \textbf{111.0(9.8)} \\
        \midrule
    \end{tabular}
    \caption{Comparative performance of HimNet and four widely used GNN baselines (MPNN, GCN, GAT, GIN) on MetStab, Malaria, and LMC (human, rat, mouse) tasks. All experiments were conducted under identical scaffold-based data splits. For each model, results are averaged over multiple runs with three different splits and random initializations; mean and standard deviation are reported.}
    \label{tab:model_performance}
\end{table*}

We first evaluated the performance of HimNet on eight classic benchmark datasets defined by MoleculeNet—including five binary classification tasks (BACE, BBBP, Tox21, SIDER, ClinTox) and three regression tasks (ESOL, FreeSolv, Lipophilicity)\cite{Wu2018MoleculeNet}—covering a broad range of physicochemical and bioactivity prediction scenarios. To further assess the applicability of HimNet in more advanced and realistic drug discovery contexts, we additionally included three less commonly studied but highly representative datasets: Malaria (anti‑malarial EC\textsubscript{50} regression)\cite{Gurung2020MalariaDataset}, LMC (liver microsomal clearance regression across species)\cite{Smith2019LiverClearance}, and MetStab (binary classification of metabolic stability)\cite{Hu2021MetabolicStability}. Although these three datasets are not newly curated, they have been much less explored in the literature and present challenging tasks that better reflect the complexity encountered in real‑world pharmaceutical research\cite{Shang2022FH-GNN}.

For all eleven datasets, we adopted scaffold‑based splitting (80\% training / 10\% validation / 10\% test) following the protocol in MoleculeNet\cite{Wu2018MoleculeNet}. Three independent splits were generated using different random seeds to ensure statistical robustness. Unlike random splitting, scaffold splitting groups molecules by their core scaffolds, minimizing the chance of similar chemical backbones appearing in both training and test sets, thus providing a more realistic and stringent assessment of model generalization\cite{Wu2018MoleculeNet}. For each split, we performed ten training runs with different random weight initializations, and all models were trained using the Adam optimizer for 100 epochs\cite{Kingma2014Adam}.

To ensure comprehensive and fair comparison, we selected a range of recent state‑of‑the‑art molecular graph neural network baselines, covering hierarchical/multiscale modeling, geometry or image augmentation, and fingerprint integration strategies. Specifically, the baselines include:
\begin{itemize}
\item Hierarchical/multiscale GNNs (e.g., MolCLR\cite{Chuang2020MolCLR}, HiMol\cite{Xie2021HiMol}, HimGNN\cite{Shang2022FH-GNN}), which represent information at atom, motif, and graph levels but often lack explicit and dynamic inter‑level interaction;
\item Geometry and image‑enhanced methods (e.g., GEM\cite{Dai2021GEM}, Uni‑Mol\cite{Gao2022UniMol}, ImageMol\cite{Guo2023ImageMol}), which enrich graph representations with 3D or visual features, though the integration with chemical semantics is often shallow;
\item Fingerprint‑augmented GNNs (e.g., FH‑GNN\cite{Shang2022FH-GNN}), which combine traditional molecular fingerprints with learned graph features, but typically only at the final output stage.
\end{itemize}

HimNet addresses these limitations by introducing an end‑to‑end bidirectional hierarchical message passing mechanism that allows atom, motif, and graph‑level representations to evolve jointly and interactively. Furthermore, HimNet employs an attention‑based fusion of molecular fingerprints and learned features throughout the entire architecture, not merely at the output. This unified and flexible framework enables the model to dynamically balance local chemical detail and global molecular context, thus enhancing its capacity to model complex molecular properties.

The experimental results, summarized in Tables~\ref{tab:table3} and~\ref{tab:model_performance}, show that HimNet achieves state‑of‑the‑art or highly competitive performance on the majority of tasks. For example, HimNet obtains the highest ROC‑AUC scores on BACE (0.890), BBBP (0.954), Tox21 (0.826), and ClinTox (0.950)\cite{Fawcett2006}. In regression benchmarks, HimNet achieves the lowest RMSE on ESOL (0.710) and FreeSolv (1.441)\cite{Chai2014}, demonstrating its strong capability in modeling physicochemical properties.

A dataset‑wise analysis reveals that HimNet also performs robustly on the less commonly studied Malaria (RMSE 1.029) and MetStab (AUC 0.896) datasets, outperforming all baselines and highlighting its generalization strength in new and challenging tasks\cite{Gurung2020MalariaDataset}. On the LMC dataset, HimNet achieves the lowest average RMSE (111.0) across the three species, with particularly strong results for human (91.7) and rat (109.3)\cite{Smith2019LiverClearance}. However, on the mouse subset (LMC‑M), the model’s performance (RMSE 132.1) is slightly behind that of GAT (124.8), likely due to higher experimental noise and inter‑species variability, which reduce the advantage of hierarchical modeling and make simpler models more robust to noise\cite{Shang2022FH-GNN}.

On the SIDER dataset, which involves multi‑label classification with highly imbalanced and long‑tailed label distribution, HimNet achieves an AUC of 0.651. Although competitive, this does not surpass some baselines; the probable reason is that SIDER contains numerous rare adverse reaction labels, making it difficult for hierarchical models to learn effective signals, while approaches that rely more on global or late‑stage feature fusion (such as FH‑GNN) may better exploit statistical correlations in imbalanced data\cite{Shang2022FH-GNN}.

For Lipophilicity regression, HimNet reaches an RMSE of 0.698, which is close to but does not exceed the best baseline performance (HimGNN at 0.632). This may be attributed to the limited sample size and the nuanced nature of lipophilicity, which depends on both local functional groups and global molecular context. While HimNet’s hierarchical approach excels when sufficient training data is available to capture these dependencies, its effectiveness is somewhat reduced in low‑data and context‑sensitive scenarios.

In summary, HimNet delivers state‑of‑the‑art or highly competitive performance on almost all molecular property prediction tasks, especially excelling on those where both hierarchical structure and chemical priors are crucial. Its relatively weaker results on datasets characterized by extreme label imbalance, small sample size, or high experimental noise (such as SIDER, Lipophilicity, and LMC‑M) indicate the challenges that remain for hierarchical modeling in such settings. Future work could incorporate strategies such as sample rebalancing\cite{Chawla2002SMOTE} or noise‑aware training to further improve robustness and adaptability.


\subsection{Explanation prediction}\label{subsec4}

Understanding how molecular structure influences predictive attention is essential for interpreting the decision-making process of graph-based models\cite{Vaswani2017Attention}. To this end, we conducted a case study using the BBBP dataset, which contains annotated information on the blood–brain barrier permeability of a wide range of compounds\cite{Wu2018MoleculeNet}. This property is particularly relevant in the early stages of central nervous system (CNS) drug development, where identifying molecules capable of penetrating the barrier is critical for reducing potential toxicity and improving therapeutic efficacy\cite{Pardridge2005}.

\begin{figure*}[htbp]
    \centering
    \includegraphics[width=0.9\linewidth]{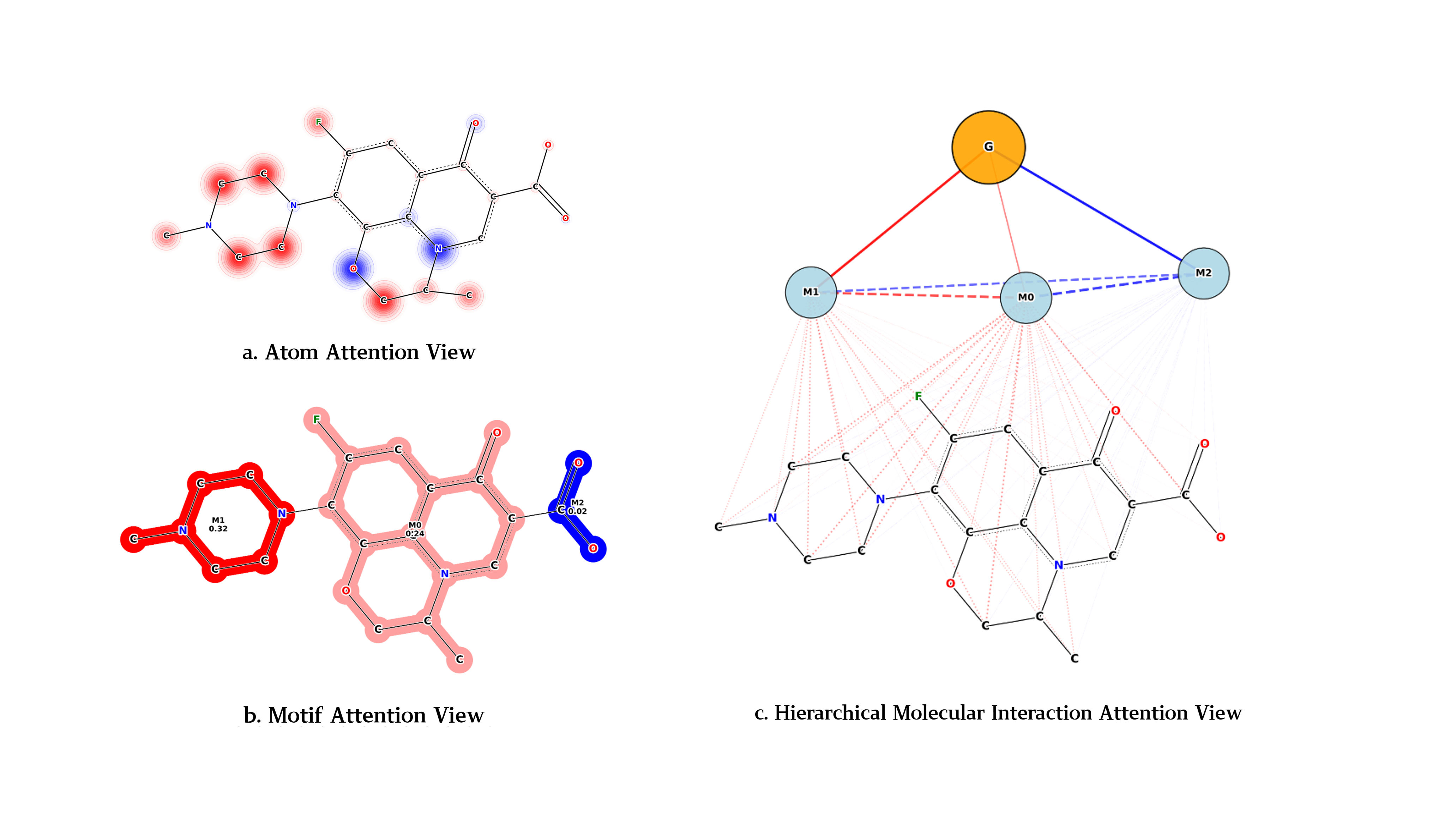}
    \caption{
    Hierarchical attention visualization for molecule CC1COc2c(N3CCN(C)CC3)c(F)cc3c(=O)c(C(=O)O)cn1c23.
    \textbf{(a) Atomic-level attention:} Highlights atom-wise contributions to BBB permeability, with red indicating positive and blue negative effects, consistent with chemical intuition that hydrophobic atoms promote and polar atoms inhibit penetration.
    \textbf{(b) Motif-level attention:} Aggregates attention to chemically meaningful substructures, showing enhanced positive attention on aromatic and fluorinated motifs, and negative attention on polar or acidic fragments.
    \textbf{(c) Hierarchical interaction attention:} Integrates atomic, motif, and global graph levels, revealing context-dependent interactions—such as competitive and cooperative effects—between motifs and their surrounding atomic environments, thereby supporting chemically interpretable property prediction.
    }
    \label{fig:visual1}
\end{figure*}

\begin{figure*}[htbp]
    \centering
    \includegraphics[width=0.7\linewidth]{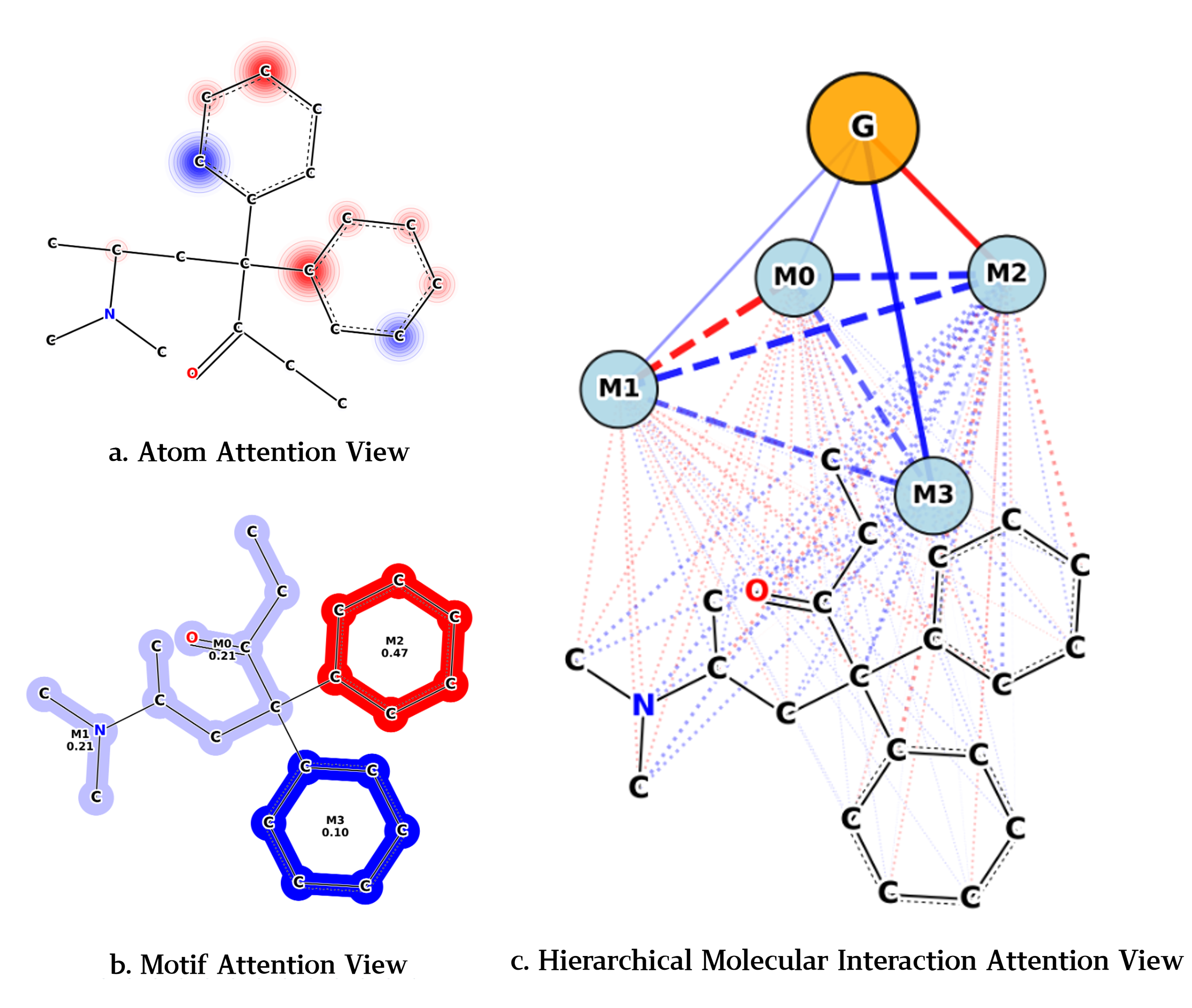}
    \caption{
    Hierarchical attention visualization of CCC(=O)C(CC(C)N(C)C)(c1ccccc1)(c1ccccc1), illustrating chemically interpretable multi-level feature interactions for BBB permeability prediction.
    \textbf{(a) Atomic-level attention:} The model assigns strong positive attention to the carbon atoms of both phenyl rings and the adjacent aliphatic carbons, reflecting their role in enhancing hydrophobicity and promoting BBB permeability. In contrast, negative attention is focused on the carbonyl oxygen and the nitrogen atom of the tertiary amine side chain, indicating these polar centers hinder BBB penetration.
    \textbf{(b) Motif-level attention:} The two phenyl rings, though structurally similar, are distinguished by the model: one is assigned strong positive attention, while the other receives negative attention, reflecting their distinct chemical environments and contextual influence from neighboring groups. The carbonyl and amine motifs both receive negative attention, consistent with their polar and hydrogen-bonding character.
    \textbf{(c) Hierarchical interaction attention:} The integrated view demonstrates that the model captures competitive and complementary relationships among motifs. The strongly positive aromatic motif (M2) competes with the negatively weighted aromatic motif (M3) and the polar carbonyl and amine motifs (M0, M1) in determining the global property. Notably, the context-dependent differentiation of the two phenyl motifs highlights the model's ability to capture non-additive and position-sensitive effects, aligning with chemical intuition regarding the interplay between hydrophobic and polar regions in BBB permeability.
    }
    \label{fig:visual2}
\end{figure*}

The visualizations in \hyperref[fig:visual1]{Fig.~2} and \hyperref[fig:visual2]{Fig.~3} illustrate how different molecular substructures contribute to blood–brain barrier permeability as interpreted by the model's attention mechanisms\cite{Gnes2020MolInterpret}. Regions with higher attention scores—depicted as brighter areas—are considered to have a greater influence on the model’s prediction of permeability\cite{Vaswani2017Attention}. In these visualizations, red-colored regions indicate molecular components that are predicted to facilitate BBB penetration, whereas blue-colored regions correspond to structural elements that are predicted to hinder it.

\paragraph{\textbf{Analysis of Molecule 1: CC1COc2c(N3CCN(C)CC3)c(F)\\cc3c(=O)c(C(=O)O)cn1c23}}

\hyperref[fig:visual1]{Fig.~2a} displays the distribution of attention weights at the atomic level within the hierarchical molecular graph. The hue and intensity of each atom indicate both the direction and magnitude of its contribution to BBB permeability as assessed by the model. Notably, red regions are concentrated on the piperazine ring and its adjacent carbon atoms on the left side of the molecule, as well as certain segments of the fused ring skeleton. From a chemical perspective, the piperazine ring, despite containing nitrogen atoms, exhibits a more aliphatic character compared to regions such as carboxyl groups\cite{Lipinski2001}. This property increases the molecule's overall lipophilicity, which may be attributed to enhanced permeability across the lipid-rich blood–brain barrier\cite{Pardridge2005}. Thus, the model's assignment of red coloration—indicating a positive contribution—is consistent with established chemical understanding.

Conversely, blue regions are predominantly found on the carboxyl (–COOH) group at the right side of the molecule, along with nitrogen and carbonyl oxygen atoms within the fused ring structure. The carboxyl group, being highly polar, is likely to dissociate into a negatively charged carboxylate (--COO$^{-}$) at physiological pH\cite{Atkins2017}. Its strong polarity, negative charge, and capacity for hydrogen bonding substantially increase the polar surface area and reduce lipophilicity\cite{Ertl2000}, both factors known to hinder BBB penetration. Similarly, the presence of polar ring nitrogen and carbonyl oxygen atoms, which can form hydrogen bonds, further elevates the polar surface area, thereby impeding BBB permeability\cite{Ekins2002}. This suggests that the model's attention allocation closely aligns with fundamental chemical principles\cite{Gnes2020MolInterpret}.

\hyperref[fig:visual1]{Fig.~2b} aggregates atomic contributions into larger chemical motifs or functional groups, illustrating their collective attention weights. In this visualization, the red region on the left corresponds to the piperazine ring (M1), the blue region on the right represents the carboxyl group (M2), and the light red area in the center denotes the core fused ring scaffold (M0). The model thus identifies the piperazine motif as generally facilitating blood–brain barrier permeability, while the carboxyl motif acts as a barrier—findings that are consistent with the chemical rationale discussed above.

\hyperref[fig:visual1]{Fig.~2c} provides an integrated, hierarchical perspective on attention allocation within the molecular graph. This visualization reveals that the model's prediction of blood–brain barrier permeability is shaped not only by the intrinsic properties of individual motifs, but also by the patterns of interaction among them\cite{Gnes2020MolInterpret}.

Crucially, these findings directly reflect the internal working mechanism of HimNet, which leverages hierarchical message passing across atomic, motif, and graph levels to encode chemically meaningful multi-scale dependencies\cite{Xie2021HiMol,Shang2022FH-GNN}. Within HimNet, atomic-level messages are first aggregated into motif-level representations via learnable, context-aware attention\cite{Vaswani2017Attention,Gnes2020MolInterpret}, capturing the essential chemical environments of functional groups. Subsequently, motifs communicate with each other and with the graph-level node through cross-level attention pathways, allowing the model to explicitly encode the non-additive, context-dependent interactions between different structural components\cite{Vaswani2017Attention,Xie2021HiMol}.

This mechanism is fundamentally distinct from conventional node-level message passing: instead of simply summing local neighborhoods\cite{Gilmer2017MPNN}, HimNet’s hierarchical structure enables a motif to reference not only its own atoms but also other motifs and even remote atomic environments\cite{Xie2021HiMol,Shang2022FH-GNN}. For example, in this molecule, the positive effect of the fluorinated aromatic motif (M0) and the pyridine motif (M1) is not simply a result of their local features, but is contextually modulated and even amplified in the presence of the polar ester motif (M2)\cite{wu2023chemistry}. Conversely, the negative contribution of M2 is intensified when considered together with neighboring hydrophobic motifs, reflecting a chemically accurate picture where polarity and hydrophobicity are in competition\cite{Chandler2005Hydrophobic}.

This direct, multi-scale message exchange and attention aggregation empower HimNet to model higher-order cooperative and antagonistic effects (such as conjugation, hydrogen bonding networks, or lipophilic clustering)\cite{Jeffrey1997HydrogenBonding,Hunter1990Aromatic,Chandler2005Hydrophobic} that underlie complex molecular properties.

In practice, each motif node in HimNet acts both as an aggregator of atomic information and as an active agent in hierarchical interactions\cite{Xie2021HiMol}. Motif-to-motif and motif-to-atom attention edges, which can extend beyond motif boundaries, provide a mechanism for encoding the chemical “fuzziness” of functional group effects\cite{Gnes2020MolInterpret}—allowing, for example, the ester motif (M2) to reference adjacent aromatic carbons (in M0), and vice versa\cite{wu2023chemistry}. These cross-hierarchy, cross-context interactions mean that property predictions arise not from linear combinations of isolated features, but from coordinated, emergent behavior across the molecular graph, providing a rigorous theoretical foundation for chemical interpretability\cite{Gnes2020MolInterpret}.

A closer examination reveals that each motif node functions not only as an aggregator of its constituent atomic information, but also as an active participant in interactions with atoms outside its defined boundaries\cite{Vaswani2017Attention,Xie2021HiMol}. For instance, the ester motif (M2) establishes links with aromatic carbon atoms in the fluorinated core (M0), thereby enhancing its disruptive polarity in an otherwise lipophilic environment. Conversely, M0 accesses information from carbonyl oxygen atoms in M2, potentially counteracting the negative effects of excessive polarity via local contextual adjustment\cite{Xie2021HiMol}. These cross-motif and motif–atom interactions indicate that the model learns context-aware semantic representations for each structural unit, allowing it to assess the contribution of each motif within its broader chemical environment\cite{Vaswani2017Attention}. Ultimately, this demonstrates that molecular properties constitute necessary consequences of coordinated component interactions, rather than mere linear combinations of isolated motifs\cite{Gnes2020MolInterpret}.

\paragraph{\textbf{Analysis of Molecule 2: CCC(=O)C(CC(C)N(C)C)(c1cc\\ccc1)(c1ccccc1)}}

In the case of the second molecule, the atomic-level attention map displays a distinct distribution pattern. As depicted in \hyperref[fig:visual2]{Fig.~3a}, regions of high attention—marked in red—are primarily located on the carbon atoms of the two benzene rings and parts of the central carbon framework. This pattern suggests that the model identifies these hydrophobic domains as positive contributors to blood–brain barrier permeability, aligning well with established chemical knowledge that lipophilic aromatic groups tend to facilitate BBB penetration\cite{Pardridge2005}.

Conversely, areas of blue coloration are centered around the nitrogen atom in the dimethylamino group (–N(CH$_3$)$_2$) and the oxygen atom in the carbonyl group (C=O). While tertiary amines may, in some contexts, promote BBB permeability, the model here assigns a negative contribution to this functional group, which may be attributed to unfavorable interactions with neighboring structural motifs\cite{Ekins2002}. The carbonyl oxygen is also assigned negative attention, consistent with its high polarity and capacity to serve as a hydrogen bond acceptor, both of which are known to impede BBB permeability\cite{Ekins2002}.

The motif-level attention analysis in \hyperref[fig:visual2]{Fig.~3b} highlights three principal functional groups, each characterized by a distinct attention weight. The first, the dimethylamino fatty chain (M1), receives a light blue score of 0.21. The ketone motif (M0) is similarly assigned 0.21, reflecting its polar character. Among the aromatic moieties, one phenyl ring (M2) is rendered in strong red with a score of 0.47, while the other ring (M3) appears in blue with a lower score of 0.10. This clear disparity between two structurally similar benzene rings demonstrates the model’s ability to differentiate aromatic systems based on their specific molecular environments and contextual interactions, rather than treating them as equivalent substructures\cite{wu2023chemistry}.

\hyperref[fig:visual2]{Fig.~3c} provides further insight into how the HimNet model’s hierarchical message passing mechanism enables the emergence of chemically meaningful, context‐dependent interactions among structural motifs. Within this integrated view, each motif not only aggregates information from its constituent atoms but also exchanges messages with other motifs and, crucially, interacts with both adjacent and remote atomic environments across the molecule\cite{Xie2021HiMol,Vaswani2017Attention}. The attention pathways learned by HimNet reflect a multi‐scale theoretical framework, in which atom‐level messages are first pooled into motif representations via attention, and subsequently, motifs propagate their semantic information to the global graph node through additional cross‐level message passing\cite{Gilmer2017MPNN}.

A key manifestation of this mechanism is that the two phenyl motifs (M2 and M3), though structurally similar, are contextually distinguished: one receives strong positive attention, while the other is negatively weighted, depending on the influence of their spatial and electronic environment—such as proximity to polar carbonyl (M0) or amine (M1) motifs\cite{wu2023chemistry}. This non‐additive and position‐sensitive differentiation cannot be achieved by conventional graph networks relying solely on local aggregation, but arises naturally from HimNet's explicit cross‐hierarchy message flow\cite{Xie2021HiMol}.

Moreover, polar motifs (M0, M1) exert inhibitory effects not only through their intrinsic properties but also by influencing the contribution of nearby aromatic regions, mirroring competitive and cooperative interactions between hydrophobic and polar domains seen in real chemical systems\cite{Chandler2005Hydrophobic,Jeffrey1997HydrogenBonding}. Motif–atom and motif–motif interactions further illustrate the model’s capability to capture “boundary effects” and extended chemical context, reflecting inductive, conjugative, or steric effects that propagate beyond simple substructure boundaries.

The patterns observed in both molecules thus exemplify the model's ability to construct a hierarchical, context‐aware representation of molecular structure—progressing from atom‐level features to motifs, and ultimately to global property prediction\cite{Gnes2020MolInterpret}. HimNet’s layered message passing and explicit multi‐scale aggregation capture not only local chemical attributes but also their broader contextual significance, ensuring that property prediction is grounded in the structured, emergent interactions among all relevant molecular components\cite{Xie2021HiMol}.



\begin{table*}[htbp]
\centering
\resizebox{\textwidth}{!}{%
\begin{tabular}{lcccccccc}
\toprule
\textbf{Variant} & \textbf{BACE} & \textbf{BBBP} & \textbf{Tox21} & \textbf{SIDER} & \textbf{ClinTox} & \textbf{ESOL} & \textbf{Freesolv} & \textbf{Lipophilicity} \\
\midrule
HimNet & \textbf{0.876(0.002)} & \textbf{0.962(0.002)} & \textbf{0.835(0.006)} & \textbf{0.651(0.002)} & \textbf{0.983(0.002)} & \textbf{0.712(0.017)} & \textbf{1.226(0.033)} & \textbf{0.711(0.002)} \\
w/o M & 0.868(0.004) & 0.959(0.005) & 0.825(0.004) & 0.645(0.005) & 0.982(0.007) & 0.738(0.008) & 1.242(0.020) & 0.723(0.005) \\
w/o I & 0.863(0.006) & 0.958(0.002) & 0.827(0.004) & 0.636(0.004) & 0.979(0.002) & 0.768(0.012) & 1.310(0.023) & 0.728(0.007) \\
w/o C & 0.864(0.001) & 0.960(0.004) & 0.831(0.008) & 0.631(0.008) & 0.975(0.009) & 0.826(0.010) & 1.656(0.056) & 0.764(0.013) \\
w/o A & 0.865(0.002) & 0.952(0.004) & 0.828(0.001) & 0.647(0.002) & 0.971(0.003) & 0.783(0.019) & 1.273(0.059) & 0.717(0.002) \\
w/o H & 0.865(0.006) & 0.955(0.009) & 0.826(0.003) & 0.645(0.010) & 0.972(0.012) & 0.792(0.015) & 1.238(0.035) & 0.731(0.008) \\
\midrule
\end{tabular}
}
\caption{Ablation results for HimNet: performance of the full model and its variants with each core module removed. The best result for each dataset is highlighted in bold. For clarity, the abbreviations for the ablated variants are defined as follows. 
\textbf{w/o M}: removes the Dual message passing module; 
\textbf{w/o I}: removes the Hierarchical interaction module; 
\textbf{w/o C}: removes the Consensus fingerprint enhancement module; 
\textbf{w/o A}: removes the Attention-guided fusion module; 
\textbf{w/o H}: removes the Hierarchical message passing module.
}
\label{tab:ablation}
\end{table*}

\begin{figure*}[htbp]
\centering
\includegraphics[width=\linewidth]{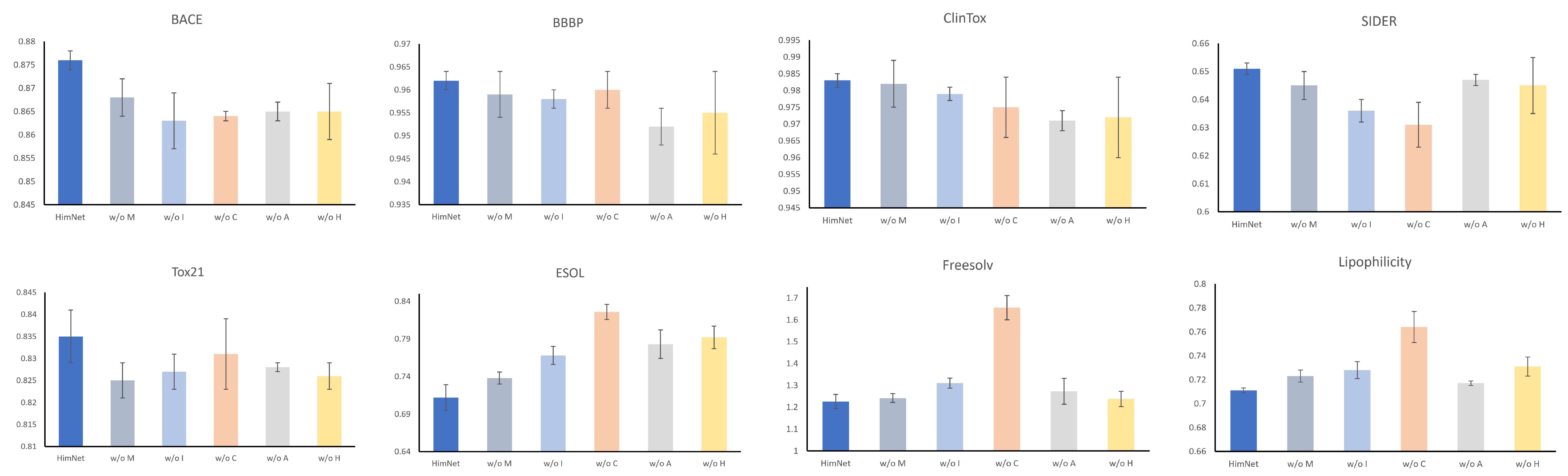}
\caption{
Performance comparison of HimNet and its ablated variants on eight benchmark datasets. Each bar shows the mean performance metric of the full model (HimNet) or a variant with a specific module removed (see legend for abbreviations). Error bars represent the standard deviation across repeated runs. Removing any module leads to a consistent performance drop, confirming the necessity of each component.
}
\label{fig:ablation}
\end{figure*}
\subsection{Ablation Study}\label{subsec5}

To rigorously assess the contribution of each module in HimNet, we conducted ablation experiments by selectively removing key components and evaluating each variant on eight widely‐used MoleculeNet benchmarks\cite{Wu2018MoleculeNet,Xie2021HiMol}. These benchmarks span both classification and regression tasks and are well‐established in the literature for systematic model analysis. The three additional datasets (Malaria, LMC, and MetStab), introduced in our main performance comparison to further demonstrate HimNet’s generalizability on challenging real‐world endpoints, are not included in the ablation study. This choice ensures clarity and interpretability of the ablation results, as the selected benchmarks already provide a comprehensive and representative testbed for isolating the effect of each architectural component\cite{Shang2022FH-GNN}. The ablation settings are as follows:


A summary of results is presented in Table~\ref{tab:ablation} and visualized in Figure~\ref{fig:ablation}, where each row (table) and colored bar (figure) corresponds directly to a specific model variant.

Table~\ref{tab:ablation} quantitatively reports the mean and standard deviation of performance metrics (e.g., ROC‐AUC for classification, RMSE for regression) across three runs on each benchmark dataset\cite{Wu2018MoleculeNet}. Figure~\ref{fig:ablation} provides a visual comparison, in which the performance of each ablated variant is shown alongside the full model. This enables a direct and intuitive understanding of the impact of each component.

Across all datasets, the full HimNet model consistently achieves the best or near‐best results (bolded in the table, leftmost bars in the figure). Removing any individual module results in a noticeable performance drop across most tasks, as clearly seen in both the numerical values in Table~\ref{tab:ablation} and the relative bar heights in Figure~\ref{fig:ablation}. In particular, removing the global‐local attention (w/o I) or the consensus fingerprint enhancement (w/o C) leads to especially large decreases in both classification (e.g., Tox21, SIDER) and regression (e.g., FreeSolv, ESOL) datasets. This is evident in Table~\ref{tab:ablation} by the lower scores in these columns, and in Figure~\ref{fig:ablation} where the corresponding bars dip most significantly below the full model\cite{Xie2021HiMol}. The necessity of multi‐head attention fusion (w/o A) is most apparent in tasks such as ClinTox and Lipophilicity, where its removal causes a marked reduction in performance (also highlighted in both the table and figure). Even the core hierarchical message passing modules (w/o M and w/o H) remain foundational, as their ablation consistently results in less accurate predictions across all datasets.

In summary, the ablation results—numerically detailed in Table~\ref{tab:ablation} and visually corroborated by Figure~\ref{fig:ablation}—demonstrate that each module in HimNet plays a vital and complementary role. The systematic drop in performance upon any module’s removal validates that these architectural components are indispensable for robust and accurate molecular property prediction\cite{Shang2022FH-GNN}.

\section{Conclusion}

In this work, we have addressed the critical challenge of accurately predicting small-molecule properties—particularly ADMET profiles—in modern drug discovery by introducing a \emph{Hierarchical Interaction Message Passing Mechanism} at the core of our novel model, \textbf{HimNet}. Unlike existing deep learning approaches (e.g., GNNs\cite{gilmer2017neural} and Transformers\cite{Vaswani2017Attention}) that treat molecular graphs in a largely flat or fragment-centric manner, HimNet explicitly models atoms, motifs, and whole-molecule fingerprints as interlinked semantic layers. Through hierarchical attention-guided message passing, our framework enables dynamic, interaction-aware representation learning that balances global topology and local chemical environments.

Extensive evaluation on benchmark datasets—including classic tasks such as BBBP, Tox21, and ESOL\cite{Wu2018MoleculeNet}, as well as three challenging, high-value datasets for metabolic stability (MetStab)\cite{Hu2021MetabolicStability}, malaria activity (Malaria)\cite{Gurung2020MalariaDataset}, and liver microsomal clearance (LMC)\cite{Smith2019LiverClearance}—demonstrates that HimNet achieves state-of-the-art or near–state-of-the-art performance across diverse molecular property prediction tasks. Moreover, the built-in Explainable Attention Interaction\cite{Gnes2020MolInterpret} and Consensus Fingerprint Enhancement modules grant HimNet a strong degree of hierarchical interpretability, aligning model attributions with established chemical intuition on representative compounds. These results confirm that explicitly capturing cross-layer functional-group synergies is key to improving both predictive accuracy and mechanistic insight.

Despite these advances, several avenues remain for further improvement. First, extending HimNet to incorporate three-dimensional conformational data and solvent effects (e.g., as in Uni-Mol\cite{Gao2022UniMol}) may enhance its ability to predict stereospecific and environment-sensitive properties. Second, reducing the computational overhead of multi-layer attention mechanisms will be essential for large-scale virtual screening campaigns. Finally, integrating our hierarchical paradigm into multi-task or multi-omics frameworks could unlock deeper insights into polypharmacology and mechanism-of-action inference.

In summary, HimNet offers an accurate, efficient and interpretable solution for molecular activity and ADMET prediction, thereby contributing significantly to informed decision-making in the early stages of drug discovery. Future work will focus on scaling and generalizing this hierarchical interaction approach to meet the evolving demands of small-molecule design and optimization.

\section{Discussion}\label{sec5}

\subsection{Research Findings and Contributions}

This study presents HimNet, a hierarchical molecular graph neural network grounded in the principle of hierarchical interaction learning\cite{Xie2021HiMol}. By explicitly modeling interactions across atomic, motif, and molecular graph levels, HimNet captures non-additive synergistic effects between functional groups, effects often neglected by conventional GNN architectures\cite{Shang2022FH-GNN}. This multiscale interaction mechanism improves the model's ability to represent complex structure-property relationships, thus improving both predictive precision and interpretability, as confirmed by benchmark evaluations (Section~\ref{sec3}, Tables~\ref{tab:table3}, \ref{tab:model_performance}).

A central innovation of HimNet lies in its dual‐pathway message passing architecture, which, in conjunction with multi‐level attention, enables the encoding of cross‐scale, nonlinear dependencies and functional group cooperativity—critical features for accurate molecular property prediction\cite{Vaswani2017Attention}. Complementing this architecture, a consensus fingerprint fusion module is introduced to adaptively extract and integrate shared chemical information from diverse fingerprint types\cite{rogers2010extended}. This enrichment of molecular representation facilitates alignment between global and local semantic features. Furthermore, HimNet leverages multi‐head attention to align fingerprint‐based and graph‐based representations at different levels, promoting both generalization and interpretability in its predictive outputs.

Empirical results demonstrate that HimNet consistently outperforms established baselines in accuracy and generalizability\cite{Wu2018MoleculeNet}. Beyond performance metrics, ablation experiments and visualization analyses highlight that explicitly capturing hierarchical cooperative effects leads not only to improved predictive power but also to chemically meaningful interpretability of model behavior\cite{Gnes2020MolInterpret}.

\subsection{Applicability and Limitations}

HimNet offers a general and effective framework for modeling multi‐scale molecular interactions, with particular advantages in scenarios where molecular properties are influenced by the cooperative behavior of substructures—such as in small molecules, drug‐like compounds, and ligands of moderate topological complexity.

Nonetheless, the framework exhibits limitations in handling molecules with high conformational flexibility, complex stereochemistry, or macrocyclic structures. In such cases, the current BRICS‐based decomposition\cite{degen2008automated} and two‐dimensional graph representation\cite{Weininger1988SMILES} may fall short, as they do not fully capture essential three‐dimensional conformational effects or dynamic interactions. Properties like conformational entropy, chiral recognition, and shape‐dependent permeability often depend on these spatial characteristics, underscoring the need for incorporating 3D structural information or alternative decomposition strategies in future developments.

Moreover, while attention mechanisms in HimNet provide a degree of interpretability by highlighting influential atoms or motifs, it is important to acknowledge their limitations: attention weights, although informative, do not inherently reflect causal relationships and should be interpreted as heuristic guides rather than definitive indicators of feature importance.

\subsection{Addressing Limitations}

To mitigate the outlined limitations, several practical improvements can be pursued. First, enhancing computational efficiency through model compression or streamlined attention mechanisms would enable scaling HimNet to larger datasets without compromising performance. Second, data augmentation and robustness techniques, such as SMILES enumeration, graph perturbations, or self‐supervised pretraining\cite{Chuang2020MolCLR}, can improve generalization, particularly in settings with limited or imbalanced data. Third, to bolster interpretability, attention‐based methods can be complemented with more rigorous explainable AI approaches—such as Shapley value analysis or counterfactual explanations—which provide deeper insights into model decision‐making\cite{Gnes2020MolInterpret}.

\subsection{Future Research Directions}

Building on the current work, several avenues merit further investigation. Integrating 3D geometric and stereochemical information into the hierarchical representation—as demonstrated in Uni‑Mol\cite{Gao2022UniMol}—would extend the model’s applicability to flexible and chiral molecules. Additionally, scaling HimNet to support large‐scale datasets and high‐throughput screening applications, or adapting it for transfer learning in related domains (e.g., protein–ligand interaction prediction or materials informatics), offers substantial promise. Finally, incorporating multimodal molecular data, including protein structures, experimental spectra, or molecular dynamics trajectories, could establish a comprehensive framework for property prediction that captures the full complexity of real‑world chemical systems.

\section{Author contributions}
Huiyang Hong conceived and designed the study, performed experiments, and drafted the manuscript. Xinkai Wu assisted with experiments and manuscript writing. Hongyu Sun assisted with experiments and revised the model code. Chaoyang Xie contributed to the initial model design and discussions on innovative ideas. Yuquan Li and Qi Wang supervised the project and contributed to all aspects of the research and manuscript.

\section{Funding}
This research is supported by Guizhou Provincial Science and Technology Projects (No.CXTD[2023]027) and the Guiyang Guian Science and Technology Talent Training Project (No.[2024]2-15).

\section{Data and Code availability}
Source code and all datasets used in this study are available at \href{https://github.com/Hugh415/HimNet}{https://github.com/Hugh415/HimNet}.

\bibliographystyle{unsrt}
\bibliography{reference}

\end{document}